
\magnification 1200
\baselineskip = 24pt
\centerline{\bf Trapped surfaces and spherical closed cosmologies.}
\vskip 2cm
\centerline{Edward Malec$^{+}$ and Niall \'O Murchadha$^*$           }
\vskip 2cm
\centerline{$^{+}$ Instytut Fizyki, Uniwersytet Jagiello\'nski
30-064 Krak\'ow, Reymonta 4, Poland }
\centerline{$^*$ Physics Department, University College,
Cork, Ireland}
\vskip 2cm
\centerline{PACS numbers: 04.20., 98.80., 95.30., 97.60.}
\vskip 0.5cm
\centerline{\bf Abstract.}
This article gives necessary and sufficient conditions for the formation of
trapped surfaces in spherically symmetric initial data defined on a closed
manifold. Such  trapped surfaces surround a region in which there occurs an
enhancement of matter over the average. The conditions are posed directly in
terms of physical variables and show that what one needs is a relatively large
amount of excess matter confined to a small volume. The expansion of the
universe and an outward flow of matter oppose the formation of trapped
surfaces;
an inward flow of matter helps. The model can be regarded as a
Friedmann-Lema\^\i tre-Walker cosmology with localized spherical
inhomogeneities
   .
We show that the total excess mass cannot be too large.\vfill\eject

\centerline{\bf I Introduction}
\medskip
We intend to study the geometry of initial data for the Einstein equations
coupled with a matter field so as to investigate the presence (or absence) of
trapped surfaces. A trapped surface is a 2-surface which exists at a particular
instant of time and has the property that all the outgoing light-rays from it
converge [1,2]. One expects that outgoing light-rays diverge, so one can
immediately deduce that the gravitational field in the vicinity of a trapped
surface must be very strong so as to prevent the light-rays from expanding. The
presence of a trapped surface indicates that the spacetime is undergoing
gravitational collapse. One of the singularity theorems of general relativity
states that if one has a trapped surface (and if the enclosed volume is finite)
then there must be a singularity to the future
(e.g., [2]). If one accepts Cosmic
Censorship, that singularities are hidden, then the presence of a trapped
surface is a sign that a black hole is in the process of forming.

Initial data for the gravitational field consists of four objects $[g_{ab},
K^{ab}, \rho \hbox{ and }J^a]$, where $g_{ab}$ is the three-dimensional
positive
definite metric of a 3-hypersurface $\Sigma$;\ $\Sigma$ is to be regarded as a
spacelike slice through the 4-manifold. $K^{ab}$ is a symmetric 3-tensor which
is the extrinsic curvature of $\Sigma$ as an embedded surface, $\rho$ is the
energy density and $J^a$ is the momentum density of the matter. These data are
not independent, they must satisfy the constraint equations
$$
^{(3)}R[g] - K_{ab}K^{ab} +(K^{a}{}_{a})^2 = 16\pi \rho\;, \eqno(1)
$$
$$
\nabla_a K^{ab} - \nabla^b K^{a}{}_{a} = -8 \pi J^b\;, \eqno(2)
$$
where $^{(3)}R[g]$ is the scalar curvature of $\Sigma$ and $\nabla$ is the
3-covariant derivative compatible with $g_{ab}$.

The trace of the extrinsic curvature ($trK = g_{ab}K^{ab}$) is equal to the
(positive) time rate of change of the three-volume, ${d \over dt}(d^3V) =
trK\,dV$. It is very common to place conditions on trK so as to select a
preferential slicing (``extrinsic time''). In asymptotically flat space-times a
standard choice is the maximal slicing condition, trK = 0. In cosmological
models(e.g. [3]), the analogous condition is to choose trK = constant.

A trapped surface is defined
as a compact two-dimensional (smooth) spacelike surface S having the property
that the expansion $\theta$ of outgoing future-directed null geodesics which
are
orthogonal to S is everywhere negative on S. If S is regarded as a submanifold
of $\Sigma$, then $\theta$ can be expressed in terms of $g_{ab}$ and $K^{ab}$
by
 $$  \theta =  \nabla_an^a - K_{ab}n^an^b + g_{ab}K^{ab}\;,\eqno(3) $$
 where
$n^a$ is the outward unit normal to S in $\Sigma$.

We have recently derived simple necessary and sufficient conditions for the
appearance of trapped surfaces in asymptotically flat initial data sets [4] and
in open universes [5], with the additional assumption that the initial data are
spherically symmetric. In this article we will derive equivalent
necessary and sufficient conditions for trapped surfaces in  a closed universe.
We continue to demand spherical symmetry and we impose the
condition that the data satisfy trK = constant.
The problem we wish to consider looks like a spherical lump superimposed on a
standard Friedmann-Lema\^\i tre-Walker (FLW hereafter) background, in
the fact that we assume that the density  becomes constant outside some
finite subset  and the matter current vanishes outside the same subset. We
address also the situation where a closed universe is filled with many
spherically symmetric bumps that have a small domain of influence so that in
between them geometry coincides with the  homogeneous and isotropic FLW
geometry
   .

The standard time-slice through a closed FLW cosmology is defined by a
three-manifold which is just the standard round three-sphere, i.e., the
line-element can be written as
$$ds^2 = a^2[dr^2 + sin^2 r d\Omega^2]\;, \eqno(4) $$
where $a$ is a constant; $K^{ab}$ is pure trace, i.e., $K^{ab} =
H g^{ab}$, where $H$ is a constant, the Hubble constant; the energy density
is a constant, $\rho_0$, and the matter current density is zero. The scalar
curvature of the manifold is given by
$$
^{(3)}R = {6 \over a^2}\;,  \eqno(5) $$
so therefore the Hamiltonian constraint reduces to
$$ {6 \over a^2} + 6H^2 = 16\pi\rho_0\;.     \eqno(6) $$
The momentum constraint is trivial.

The normal $n^a$ to the two-surfaces of constant radius is given by
$ n^a = (a^{-1 }, 0, 0)$ and its divergence is given by
$$ \nabla_a n^a = (a\sin^2 r)^{-1 }{d\sin^2 r
\over dr} = {2 \over a}\cot r\;. \eqno(7) $$
Therefore the expansion $\theta$ of such surfaces is given by
$$ \theta = {2\cot r + 2aH \over a}\;. \eqno(8)$$
If the universe is expanding ($H > 0$), then trapped surfaces can only occur on
the `other' side of the equator, i.e., $r > {\pi \over 2}$, where $\cot r$ is
negative. Since $\cot r$ becomes unboundedly large and negative, we will always
find trapped surfaces. These trapped surfaces indicate the existence of the
final `big crunch'. This property should hold for any spherically symmetric
universe and so finding trapped surfaces near the `south pole' is not very
interesting. Thus we will be interested in finding trapped surfaces in the
 `northern' hemisphere, far away from the `south pole'. These, we expect,
should
indicate the onset of some local gravitational collapse, superimposed on the
overall expansion and subsequent collapse of the whole universe.

 The initial data $[g_{ab},K^{ab}, \rho, J^a]$ we wish to consider
are defined on a three-manifold $\Sigma$, which has topology $S^3$. We assume
$trK = g_{ab}K^{ab} = 3H = $ constant holds everywhere on the 3-manifold. We
further assume that the metric, the extrinsic curvature, $\rho$,
and $J^a$ are spherically symmetric. Because of the spherical symmetry we
know that the three-metric is conformally flat and so can write the
line-element in isotropic coordinates as
 $$ds^2 = a^2\phi^4[dr^2 + sin^2 r d\Omega^2]\;, \eqno(9) $$
where $a$ is a positive constant and $\phi$ is a positive conformal factor that
depends only on $r$. We assume that $\rho$ and $J^a$ are arbitrary (other
than satisfying some positivity conditions one may wish to impose). We do not
impose any equation of state on the matter. We need not do so as we restrict
attention to a single instant of time, we make no use of the Einstein evolution
equations. We further assume that $\rho = \rho_0$ (a constant) and $J^a$ = 0
outside some subregion of compact support. We choose the constant $a$ in
eqn.(9)
to satisfy

 $$ {6 \over
a^2} + 6H^2 = 16\pi\rho_0     \eqno(10) $$
so as emphasise the link with the FLW cosmology.

\vskip 0.5cm
\centerline{\bf II The extrinsic curvature in spherically symmetric
closed universes}
\medskip
 We assume that the extrinsic curvature is spherically symmetric. This means
that the general form of $K^{ab}$ is
$$ K^{ab} = K(r)n^an^b + b(r)g^{ab}\;, \eqno(11) $$
where $n^a$ is the unit vector in the radial direction, and $K(r)$ and $b(r)$
are two scalar functions. Since we know trK = 3H, we can rewrite (11) as
  $$ K^{ab} = Hg^{ab} + K(r)[n^an^b - {g^{ab} \over 3}]\;.
\eqno(12) $$ The momentum constraint, Eq. (2), reads
$${2 \over 3a^2\phi^4}[K' + (6{\phi' \over \phi} +3{\cos r \over \sin r})K]
   = - {8\pi J \over a\phi^2}\;, \eqno(13) $$
where we assume $J^a = Jn^a$, with $n^a = (a^{-1}\phi^{-2}, 0, 0)$ as the unit
vector in the radial direction, $J = J(r)$ is a scalar and the primes represent
derivatives with respect to $r$. Eq. (13) can be slightly simplified to
$$K' + (6{\phi' \over \phi} +3{\cos r \over \sin r})K = -12\pi a \phi^2 J\;.
\eqno(14) $$

Now let us make our standard assumption, that J is zero outside some finite
subset. The momentum constraint (Eq. (14)) reduces to
$$K' + (6{\phi' \over \phi} +3{\cos r \over \sin r})K = 0\;, \eqno(15)$$
 in the exterior region. It is easy to solve this equation to give
 $$ K = C(\phi^6 \sin^3 r)^{-1}\;, \eqno(16) $$ where $C$ is some constant.
This is a well-known result, in disguise. $K$ generates a divergence-free,
tracefree (TT) tensor. The spherically symmetric closed manifold is
conformally flat, and TT tensors are conformally covariant. There is a unique
spherically symmetric TT tensor on flat space, and the tensor $K(n^an^b
-g^{ab}/3)$, with $K$ given by Eq. (16), is exactly this tensor,
transformed in the correct way.

 The constant $C$ is essentially the integral
of the matter current. Eq. (14) can be rewritten as
 $$[K \phi^6 \sin^3 r]' =
-12 \pi a \phi^{8} \sin^3 r J\;. \eqno(17) $$
This can be integrated  from $r= 0$ to $r = R$ to give
$$K(R) = -12 \pi
a[\phi^6(R) \sin^3 R]^{-1} \int_0^R \phi^{8} \sin^3 r J dr \;. \eqno(18) $$

One important consequence of (18) is that if the matter is at rest, i.e.,
$J \equiv 0$, then we can immediately deduce that $K \equiv 0$ and the
extrinsic curvature is pure trace. Another consequence is that is that if
we have a globally spherically symmetric model with the matter current
confined to a region near the north pole, the integral in Eq. (18) when we let
R
approach $\pi$ must go to zero. Otherwise, the extrinsic curvature term in the
Hamiltonian constraint will have a term that diverges like $\sin^{-6}r$ as
r approaches $\pi$. In other words, if $J$ vanishes outside a finite
region (and we have global spherical symmetry) so also will $K$.

\vskip 0.5cm
\centerline{\bf III The Hamiltonian constraint in spherically symmetric
closed universes}
\medskip
 We now can put Eqs. (9) and (12) together and evaluate the
Hamiltonian constraint. We can write the scalar curvature as $ ^{(3)}R = 6
 a^{-2}\phi^{-4} -  8\phi^{-5}\nabla^2\phi$, where the laplacian is with
respect to the background metric (Eq. (4)). We also have that $(trK)^2 -
K^{ab}K_{ab} = 6H^2 - {2 \over 3}K^2$. Thus the hamiltonian constraint, Eq.
(1), reads

$$ {6 \over a^2}\phi^{-4} - 8\phi^{-5}\nabla^2\phi + 6H^2 - {2 \over
3}K^2 = 16 \pi \rho\;. \eqno(19) $$
Rearranging, and using Eq. (10), we get
$$\nabla^2\phi = {3 \over 4a^2}(\phi - \phi^5) - [{1 \over 12}K^2
+ 2\pi(\rho - \rho_0)]\phi^5 \;. \eqno(20) $$
This is our key equation and much of our effort will be devoted to analysing
it.

The other important quantity we need to consider is the expansion $\theta$
(Eq. (3)) of the surfaces of constant $r$. The normal to such surfaces is given
by  $n^a = (a^{-1}\phi^{-2}, 0, 0)$, and $\nabla_a n^a$ is given by
$$ \nabla_a n^a = (a\phi^6 \sin^2 r)^{-1}(\phi^4 \sin^2 r)' =
2(a \phi^3 \sin r)^{-1}(2 \phi' \sin r  + \phi \cos r)\;. \eqno(21) $$
On using Eq. (12) we get $- K_{ab}n^an^b + g_{ab}K^{ab} = 2H - {2 \over 3}K$.
Hence we get (from (3))
$$\theta = 2(a \phi^3 \sin r)^{-1}(2 \phi'\sin r  + \phi\cos r ) +
 2H - {2 \over 3}K\;. \eqno(22) $$
The aim of this article is to take the expression (22) for the expansivity in
terms of the `potentials' $\phi$ and K, and use equations (18) and (20) to
replace these quantities with $\rho$ and $\vec{J}$, more physical
objects. Two other spherically symmetric cases have already been
successfully dealt with in this fashion, the case where the three-geometry is
asymptotically flat and the case where the system asymptotically approaches a
flat Friedmann cosmology [4,5].
\vfill \eject

\centerline{\bf IV. A sufficient condition for trapped
surfaces when J = 0.}
\medskip
We assume that we are given spherically symmetric initial data  $[g_{ab},
K^{ab}, \rho, \vec{J}\;]$ (which may be confined to a finite region). We assume
that the trace of the extrinsic curvature is constant. We further specialize by
assuming that the matter-current is instantaneously zero. As we have shown in
Section II, this allows us to conclude that the extrinsic curvature is pure
trace, so that the function $K(r) \equiv 0$. We also assume that $\rho$ assumes
   a
constant value $\rho_0$ outside a region of compact support. The Hamiltonian
constraint now reduces to a simplified version of Eq. (20)
$$\nabla^2\phi = {3 \over 4a^2}(\phi - \phi^5) -2\pi\Delta\rho\phi^5 \;,
\eqno(23) $$
where $\Delta\rho = \rho - \rho_0$.

We now wish to get some relationship between our sources (in this case
$\Delta\rho$) and the expansion $\theta$ of any given spherical surface,
with coordinate radius $r = R_0$ in the northern hemisphere. The trick is to
take Eq. (23), multiply it by $\phi$ and integrate it in the background metric,
over the volume V enclosed by the surface at $R_0$. This gives
$$\int_V\phi\nabla^2\phi dV = {3 \over 4a^2}\int_V(\phi^2 - \phi^6)dV
 -2\pi\int_V\Delta\rho\phi^6 dV\;.\eqno(24)$$
The left-hand-side of (24) can now be written as
$$\eqalignno{
\int_V\phi\nabla^2\phi dV &= \oint_{R_0}\phi\nabla_a\phi\cdot dS^a
           -\int_V (\nabla\phi)^2 dV \cr
&= 4\pi a \phi \phi' \sin^2 r|_{R_0} - 4\pi a\int_0^{R_0}(\phi')^2 \sin^2 r
dr \;. &(25)\cr}$$
Now consider the expression (22) for $\theta$, dropping the term in K. This
can be rearranged to give
$$(\pi a^2 \phi^4 \sin^2r)(\theta - 2H) = 4 \pi a \phi \phi' \sin^2 r
                           + 2\pi a \phi^2 \sin r \cos r \;. \eqno(26)$$
We should recognise that $4\pi a^2 \phi^4 \sin^2r = A$ is the proper area of
a sphere of coordinate radius r in the physical space. We now merge (26) with
(25) to give
$$\int_V\phi\nabla^2\phi dV = {A \over 4}(\theta|_{R_0} - 2H)- 4\pi
a\int_0^{R_0} (\phi')^2 \sin^2 rdr -  2\pi a \phi^2 \sin R_0 \cos R_0 \;.
\eqno(27)$$
We can simplify the right-hand-side of (24) when we recognise that
$$\int_V \phi^6 dV = V \;, \eqno(28)$$
where V is the proper volume enclosed by the surface, and
$$\int_V\Delta\rho\phi^6 dV = \Delta M \;,\eqno(29)$$
where $\Delta M$ is the mass excess as measured in the physical space. We now
combine (24) and (27) to give $${A \over 4}\theta|_{R_0} = \pi
a\int_0^{R_0}[3\phi^2\sin^2 r + 4(\phi')^2 \sin^2 r + 2 (\phi^2 \sin r\cos r)']
dr + {AH \over 2} - {3V \over 4 a^2} - 2\pi\Delta M \;. \eqno(30)$$
Consider the integrand of the integral on the right-hand-side of (30). It is
$$\eqalignno{
I &= \phi^2\sin^2 r + 4(\phi')^2 \sin^2 r + 4\phi\phi'\sin r\cos r
      + 2\phi^2\cos^2 r & (31)\cr
  &= 2\phi^2 - \phi^2\sin^2 r + 4\phi'\sin r(\phi'\sin r + \phi \cos
r) \;.          &(32)\cr}$$
 Let us now make two (as yet unjustified) assumptions.
We will assume $$ \phi' < 0 \hbox{ and } \phi'\sin r + \phi \cos r > 0
\eqno(33)$$
 on the support of the matter.In later sections we will show that these
conditions follow naturally from the Hamiltonian constraint (20), if we assume
that the excess mass $\Delta \rho$ is positive. Almost identical conditions
have been derived and used in [4] and [5].
         We now can deduce
$$I < 2\phi^2 \;. \eqno(34)$$
 Thus we obtain the inequality we want
$${A \over 4}\theta|_{R_0} < 2 \pi a \int_0^{R_0}\phi^2 dr + {AH \over 2}
- {3V \over 4 a^2} - 2\pi\Delta M \;. \eqno(35)$$
We can identify  $a \int_0^{R_0}\phi^2 dr = L$ as the proper radius of the
sphere of coordinate radius $R_0$. Thus we can write inequality (35) as
$${A \over 8\pi}\theta|_{R_0} < L + {AH \over 4\pi} -{3V \over 8 \pi a^2}
     - \Delta M \;. \eqno(36) $$
We have proven the following:
\medskip
{\bf Lemma 1}:Assuming conditions (33), if we can find in
our physical data a spherical surface satisfying
$$\Delta M >  L + {AH \over 4\pi} -{3V \over 8 \pi a^2} \eqno(37)$$
then that surface is trapped.
\medskip
 If we were interested in minimal surfaces
rather than in trapped surfaces we can prove:
\medskip
{\bf Lemma 2}: Under the conditions stated in this section, if we can find
in our physical data a spherical surface satisfying
$${1 \over 16\pi} \Delta ^{(3)}R > L  -{3V \over 8 \pi a^2} \;,\eqno(38)$$
where $\Delta ^{(3)}R$ is the excess integrated scalar curvature, then
the manifold must contain a minimal surface.

\vskip 0.5cm
\centerline{\bf V. A necessary condition for trapped surfaces when J = 0}
\medskip
It is possible to find a necessary condition for trapped surfaces under much
weaker conditions. We need make no assumptions such as (33).  Let us return to
equation (30) and consider again the integrand (31)
$$\eqalignno{
I &= \phi^2\sin^2 r + 4(\phi')^2 \sin^2 r + 4\phi\phi'\sin r\cos r
      + 2\phi^2\cos^2 r \cr
 &= \phi^2 + [\phi \cos r + 2\phi' \sin r]^2 & (39)\cr
 &> \phi^2 \;. &(40) \cr}$$
It is interesting to note that we do not impose any conditions
on $\phi$ or $\phi'$ to derive this
inequality. When (40) is substituted back into (30) we get
$${A \over 4}\theta|_{R_0} > \pi a \int_0^{R_0}\phi^2 dr + {AH \over 2}
- {3V \over 4 a^2} - 2\pi\Delta M \;. \eqno(41)$$
Again we use $ \int_0^{R_0}a\phi^2 dr = L$, the proper radius of the sphere
to simplify (41) to
$${A \over 8\pi}\theta|_{R_0} > {L \over 2} + {AH \over 4\pi} -{3V \over 8 \pi
a^2}
     - \Delta M \;. \eqno(42) $$
Thus we have
\medskip
{\bf Theorem I}: Any spherical surface in the physical data satisfying
$$\Delta M <{L \over 2} + {AH \over 4\pi} - {3V \over 8\pi a^2} \eqno(43)$$
cannot be trapped.
\medskip
 Let us stress again that this inequality is valid for any
spherical surface in the whole universe. As a trivial application, it is
compatible with the fact that all the surfaces near the south pole are
trapped, even with no extra mass. As we approach the south pole, in the
standard Friedmann model, we get $L \sim a\pi, V \sim 2\pi^2a^3$ and $A \sim
0$. Thus the right-hand-side of (43) approaches $-\pi a/4$.

\vskip 0.5cm
\centerline{\bf VI. The effect of a matter current on trapped surface
formation}
\medskip
In Sections IV and V we have dealt with the situation where the
matter current was at rest. In this section we will consider the effects of
nonzero $\vec{J}$. We will continue to assume that both the excess matter and
the current density are confined to the northern hemisphere and that the trace
of the extrinsic curvature is a constant. The hamiltonian constraint has the
general form as given by (20)
$$\nabla^2\phi = {3 \over 4a^2}(\phi - \phi^5) - [{1 \over 12}K^2
+ 2\pi\Delta \rho ]\phi^5 \;, \eqno(44) $$
 and we see that the
effect of the nonzero current is just to add another positive term to
$\Delta \rho$.
We need to derive an equation similar to (30), but taking into account the
extra terms that arise in both the definition of the expansion and in the
Hamiltonian constraint.  Before we do so, it is useful to multiply
the momentum constraint by the unit radial vector and integrate it over a
spherical volume in the physical space. We assume that the extrinsic
curvature has the form given by (12)
 $$ K^{ab} = Hg^{ab} + K(r)[n^an^b - {g^{ab} \over 3}] \;,\eqno(45) $$
where H is a constant. We thus wish to consider
$$\int_V (n_a\nabla_bK^{ab} - n^a\nabla_a K^{b}{}_{b}) dV = - 8\pi \int_V
n_a J^a dV \;. \eqno(46)$$
 The term in H contributes nothing to the integral so we can ignore it.
Thus we can replace $K^{ab}$ in the integral with $k^{ab} =  K(r)[n^an^b -
g^{ab}/3]$. The second term on the left in (46) is zero as well. Now we
integrate by parts the remaining term on the left to give
$$An_an_bk^{ab}|_{R_0} -\int_V k^{ab}\nabla_an_b dV = - 8\pi \int_V
n_a J^a dV \;. \eqno(47)$$
This can be further simplified to read
$${2 \over 3} KA|_{R_0} + {1 \over 3}\int_V K \nabla_a n^a dV
= - 8\pi \int_V n_a J^a dV \;. \eqno(48)$$
Substituting from Eq. (21) finally gives
$${2 \over 3} KA|_{R_0} + {8\pi a \over 3}\int_0^{R_0}K a \phi^3 \sin r
(2\phi' \sin r + \phi \cos r) dr = - 8\pi \int_V n_a J^a dV \;. \eqno(49)$$
Let us now return to the Hamiltonian constraint. Repeating the manipulation
of Section V leads to an equation analogous to (30)
$${A \over 4}\theta|_{R_0} = \pi a \int_0^{R_0}[I -
{1 \over 3}(K a \phi^3 \sin r)^2] dr +{AH \over
2} - {AK \over 6} - {3V \over 4a^2} - 2\pi \Delta M \;, \eqno(50)$$
where I is exactly the integrand given in (31). Let us eliminate the $AK$
term from (50) by subtracting 1/4 of (49) from it. To simplify the notation
somewhat, we will replace $K a \phi^3 \sin r$ by $k$ and divide by $2\pi$ to
give
$${A \over 8\pi}\theta|_{R_0} ={a \over 2} \int_0^{R_0}[I - {1 \over 3}k^2
+ {2 \over 3}k(2\phi' \sin r + \phi \cos r)] dr + {AH \over 4\pi}
- {3V \over 8\pi a^2} -(\Delta M - \Delta J) \;, \eqno(51)$$
where we define $\Delta J = \int_V n_a J^a dV$.

We can write the total integrand in (51) as
$$I' = \phi^2(\sin^2 r + 7/3 \cos^2 r) + 8\phi'\sin r(2\phi' \sin r +
    \phi \cos r - k/3) - {1 \over 3}(k - 6\phi' \sin r -\phi \cos r)^2 \;.
\eqno(52)$$
The first term in the integrand $I'$ is easy to handle. It can be written as
$$I_1 = \phi^2(7/3 - 4/3 \sin^2 r) < {7 \over 3}\phi^2 \;, \eqno(53)$$
and the integral of this term is $< 7L/6$.
The middle term can be written, on using (22) as
$$\eqalignno{
I_2 &= 8 a \phi^3\phi'\sin^2 r(\theta/2 - H) &(54) \cr
    &= 4 a \theta \phi'\phi^3 \sin^2 r - 8Ha \phi^3 \phi'\sin^2 r &(55)\cr
   &= 4 a \theta \phi'\phi^3 \sin^2 r
 - 2Ha(\phi^4 \sin^2 r)'+ 4Ha \phi^4 \sin r \cos r \;. &(56) \cr}$$
If we have no trapped surfaces in the interior, we have that $\theta > 0$
and therefore the first term in (56) is obviously negative, on using
our standard inequality (33), i.e., $\phi' < 0$. The second term integrates to
give $-AH/4\pi$,  which cancels the equivalent term in (51). The only term that
we need to consider carefully is the third term in (56). The integral of this
term can be estimated as follows. Let us define
$$ \Gamma = 2Ha^2\int_0^R \phi^4 \sin r \cos r dr \;.       \eqno(57)$$
We can show that $\Gamma < HL^2$. To obtain this we need to use
$r^{-1}\sin r \cos r \leq 1$. This allows us to write
$$\Gamma < 2H \int_0^R (a r \phi^2)(a \phi^2 dr) \;. \eqno(58)     $$
If we assume $\phi' < 0$ we have that $\phi$ monotonically decreases so we know
that, at any point, $a \phi^2 r < L$, and, of course, $a \phi^2 dr = dL$. This
immediately gives us the desired result.

Let us now return to Eq. (52) and assume that there is no trapped surface
inside the radius $R_0$ we are considering. We can now write
$${A \over 8\pi}\theta|_{R_0} < {7 \over 6}L + HL^2 - {3V \over 8\pi a^2}
-(\Delta M - \Delta J) \;. \eqno(59)$$
This now will give us our desired result:
\medskip
{\bf Lemma 3}: Assuming $\phi' < 0$, if
we obtain a surface for which
$$\Delta M - \Delta J > {7 \over 6}L + HL^2 - {3V \over 8\pi a^2} \eqno(60)$$
either the surface is itself trapped or the interior contains trapped
surfaces.
\medskip
Another estimate can be derived if we are willing to use both conditions in
(33). We can write the total integrand $I'$ in (51) in a different way. We get
$$I' = \phi^2(\sin^2 r + 7/3 \cos^2 r) + { 16 \over 3}\phi'\sin r(\phi' \sin r
+
    \phi \cos r) - {1 \over 3}(k - 2\phi' \sin r -\phi \cos r)^2 \;.
\eqno(61)$$
If we accept (33) the middle term is negative, the last term is obviously
negative, and, as in (53), the first term is less than ${7 \over 3}\phi^2$.
This means that we can replace (59) with
$${A \over 8\pi}\theta|_{R_0} < {7 \over 6}L + {AH \over 4\pi} - {3V \over 8\pi
a^2} -(\Delta M - \Delta J) \;. \eqno(62)$$
This gives us
\medskip
{\bf Lemma 4}: Assuming conditions (33), if
we obtain a surface for which
$$\Delta M - \Delta J > {7 \over 6}L + {AH \over 4\pi} - {3V \over 8\pi a^2}
\eqno(63)$$
the surface is trapped.
\medskip
{\bf Note}: Lemma 4 is usually better than Lemma 3 because if we assume
$\phi' < 0$ we immediately get $A < 4 \pi L^2$.
\vfill\eject

\centerline{\bf  VII. The total excess mass must be bounded}
\medskip
There are many inequalities that
one can derive using these techniques, but one, in particular, is enlightening
i
   n
that it shows that there are very strict bounds on the total amount of excess
mass that can be placed in finite region, independent of whether this matter is
inside a horizon or not. Let us return to eqns.(23) and (24) but assume that
$J$ is nonzero. These can be written as
$$\int K^2 dV + 2\pi \Delta M =  -4\pi a \phi \phi' \sin^2 r|_{R_0} + \pi
a\int_0^{R_0} (3\phi^2\sin^2 r + 4(\phi')^2 \sin^2 r)dr - {3V \over
4a^2} \;.\eqno(64)$$
This can be rewritten as
$$\int K^2 dV + 2\pi\Delta M = -4\pi a\phi \sin r(\phi' \sin r +
\phi \cos r)|_{R_0} + $$
$$2 \pi a\phi^2 \sin r\cos r|_{R_0}  + \pi
a\int_0^{R_0}I dr - {3V \over 4a^2} \;, \eqno(65)$$
where I is the same integrand as in Eq. (31). The first term is negative
(on assuming (33)), and the integral is bounded above (from (34)) by $2\pi L$.
Thus we have
$$\int K^2 dV + 2\pi \Delta M < 2\pi a \phi^2 \sin r\cos r|_{R_0} + 2 \pi L
- {3V \over 4a^2} \;.\eqno(66)$$
$a \phi^2 \sin r = \hat R$ is the natural areal coordinate, the equivalent of
the Schwarzschild coordinate. Therefore we get
$$\Delta M <\hat R_0\cos R_0 + L - {3V \over 8\pi a^2} - {1 \over 24\pi}
\int K^2 dV.\eqno(67)$$
 Hence the mass of the inhomogeneity inside any given
sphere cannot exceed the sum of the proper radius and the areal radius. This is
a result that supports Einstein's view that ``matter cannot be concentrated
arbitrarily'' [6].

\centerline{\bf VIII. Nonhomogeneous cosmologies.}
\vskip 0.5cm
In Sections IV, V and VI we have derived necessary and sufficient conditions
for the appearance of trapped surfaces in spherical cosmologies. The only
assumptions we make are contained in equation (33). This section, and the
following ones will be devoted to showing that these conditions can be derived
as a consequence of the Hamiltonian constraint (20), assuming that the mass
excess be positive and localized.

We can conceive of two very different situations in which spherical symmetry
could be assumed in a cosmological context. The local situation would be where
the spherical symmetry holds only on a (small) patch of the data, a spherical
galaxy in a universe with many such objects. We will discuss this case first.
The other case is when the spherical symmetry holds globally, we have only one
spherical lump in the whole universe. We postpone discussion of this situation
until Section IX.

In the last few sections we
were concerned with universes containing an isolated spherically symmetric
lump. There may be a large number of such bumps, some of them with excess
energy but many with deficit energy. Direct integration of the Lichnerowicz
equation (Eq. (20)) shows that the total energy of all bumps vanishes. If the
universe on average is homogeneous and isotropic we expect that there exist
regions with positive and negative excess energy.\par
 We assume that those regions are
separated so that  in between them  the metric approaches
the Friedmann-Lema\^\i tre-Walker metric, i.e., the conformal factor
$\phi $ tends to 1 outside the perturbed region and the extrinsic curvature
vanishes (except, of course, for the Hubble constant term). We place the north
pole at the center of the given lump. Then one might be more specific and say
that $\phi $ tends to 1 before reaching the coordinate distance $r=\pi /4$
or even $r=5\pi /12$. That is not a severe restriction since,
as pointed above,
there might exist many bumps and a zone of influence of any of these
may cover only a small fraction of the whole universe. \par

Keeping this in mind, we prove the following:\par
 {\bf Lemma 5.} Assume that we are given an $r_0$, $0 < r_0 < 5\pi/12$
such that$\Delta \rho $ is nonnegative for $r \in (0, r_0)$ and that
$\phi(r_0) \ge 1$
then
$$\phi \ge 1,~~~~\phi' \le 0  ~~~~ \forall r \in (0, r_0)\eqno(68) $$

\par {\bf Proof.}
Using spherical symmetry, we may write equation (20) as
follows:

$$\nabla^2 \phi = {1 \over a^2\sin^2r}(\sin^2r \phi')' =
-2\pi \Delta \rho \phi^5 - K^2/12\phi^5 + 3/(4a^2) \phi (1-\phi^4) \;.
\eqno(69)$$ One easily checks that
$$\alpha_0 = \sup[3/(4) \phi(1-\phi^4) : 0 \le \phi \le 1 ]=
0.6/ 5^{1/4}=0.4012442 \;.$$
We will need below an  equation related to (66),
$$\nabla^2\chi = {1 \over a^2\sin^2r}(\sin^2r\chi')'= \alpha_0 \;. \eqno(70)
$$ Assume the contrary to the claim: let there exists
a solution $\phi$ of (24) such that $\phi' \ge 0$ on
an interval $(R_i^*, R_i)$.
 Let us remark that this means that $\phi < 1$ in a part of
this interval because if $\phi > 1$ on the whole interval, then the
right-hand-side of (66) is nonpositive. The maximum principle guarantees
that $\phi$ cannot have an interior minimum, which contradicts the
positivity of $\phi'$.\par
 Assume that $\phi(R_i^*)=\chi(R_i^*)$; by definition
$d/dr\phi(R_i^*)=0$. The equation (67) is solved by
$$\chi = -(\alpha_0/2 ) r \cot r +C_i \;. \eqno(71)$$
One can find that $\chi$ is increasing everywhere,
so that at the point $R_i^*$  we have $d/dr\chi >d/dr\phi$.
Under the above stated initial conditions, and recognising that $\nabla^2(\chi
- \phi) > 0$, one may conclude that everywhere on the chosen interval

$$d\phi/dr \le d\chi/dr , ~~~ \phi \le \chi \;.$$
The change $\delta \phi = \phi(R_i)- \phi(R_i^*)$
 of $\phi$ on $[R_i^*, R_i]$  is estimated from above by
 the change of $\chi $, given by $\chi(R)-\chi(R_i^*)$
 (note that both functions are equal at $R_i^*$ !).
But the change of $\chi$ on all
intervals $\bigcup_1^n (R_i^*, R_i),\ n\le \infty$
on which $\phi $ is increasing is smaller
than the  change of $\chi $ on the entire interval $(0, 5\pi /12)$,
which is not greater than $\alpha/3$. Since at that point $\phi \ge 1$,
it means that $\phi \ge 1-\alpha_0/3$.
\par
 Now, let us observe that the last term on the
right hand side of (24) is decreasing for $\phi > 1-\alpha_0/3$.
Define
 $$ \alpha_1 = \sup[ 3\phi(1-\phi^4)/4: 1 \ge \phi \ge 1- \alpha_0/3] \;,$$
according to the above  remark, $\alpha_1 < \alpha_0$. Repeating the
above described procedure, we will obtain a better estimation from
below for the function $\phi$,
$$\phi  > 1- {\alpha_1 \over 3} \;.$$
Repeating the above procedure infinitely many times, we finally
arrive at the desired estimation
$$\phi \ge  1 \;.$$
{}From Eq. (69) we see that $\phi$ is monotonically decreasing,
since the right hand side of (69) is nonpositive
and we may apply the maximum principle. That ends the proof
of the lemma. \par
{\bf Remark.} That iteration procedure that has been described
above works until one meets a fixed point of the
map:
$$ \phi_{i+1} = 1- \phi_i (1-\phi_i^4)/4 \;,$$
i. e., a value $\phi_*$ such that
$$\phi_*=1-\phi (1-\phi_*^4)/4 \;.$$
The only solution of the last equation
is $\phi_*=1$.

One situation that is of interest is where one has an isolated lump. This is
the situation where the deviation from  the Friedmann background is localized
in the sense that the conformal factor equals 1
somewhere in a region outside the lump.
Such an isolated lump may be formed by having a region of enhanced density
surrounded by a region with diminished density In such a situation we can prove
\medskip
{\bf Lemma 6}: Given three radii $r_0, r_1, r_2$ with $r_0 < r_1 < r_2$,
$r_0 < 5\pi /12$ and $r_1 < \pi /2$, such that $\Delta \rho \ge 0$ for $r \in
(0, r_0)$, $\Delta \rho \le 0$ for $r \in (r_0, r_1)$ and $\Delta \rho = 0$
for $r \in (r_1, r_2)$. If $\phi = 1$ for a particular $r \in (r_1, r_2)$ and
if
$\phi < (\sin r)^{-1/2}$ then $\phi > 1, \phi' \le 0$ for $r \in (0, r_1)$.
\medskip
We will postpone the proof of this Lemma until the end of Section X.
\medskip
Lemma 6 and Lemma 3 can now be combined to prove
\medskip
{\bf Theorem II}: Given an isolated spherical lump with a central region of
positive excess mass and if in that region we find a surface satisfying
$$\Delta M - \Delta J > {7 \over 6}L + HL^2 - {3V \over 8\pi a^2}$$
we must have a trapped surface.
\medskip
 We will also prove the following.
\medskip
{\bf Lemma 7.}If $\Delta \rho $ is
nonnegative and if $(\phi'\sin r + \phi cos r)|_{r_0} \ge 0$ then
$$\phi'\sin r + \phi cos r \ge 0, \forall r \in (0, r_0) \;. \eqno(72)$$
{\bf Proof. } Notice that
$$ {1 \over a^2\sin r} (\phi' \sin r + \phi \cos r)' =
{1 \over a^2\sin^2 r}(\sin^2 r \phi')' -\phi/a \eqno(73) $$
$$= -2\pi \Delta \rho \phi^5 - ({\phi \over 4a^2} +3{\phi^5 \over 4a^2}) -
K^2\phi^5/12 \;.\eqno(74)$$
 The last equation follows from the constraint equation
(24). The quantity $\phi'\sin r + \phi cos r$ is decreasing;
but at $r_0$ it is  $> 0$, hence
 $\phi' \sin r + \phi \cos r$
is positive everywhere for $r \in (0, r_0)$.

\vskip 0.5cm
\centerline{\bf IX Spherically symmetric constant scalar curvature manifolds}
\medskip
Until now we have been dealing with localized spherically symmetric lumps which
we regarded as being embedded in a Friedmann background. The class of
spherically symmetric cosmologies contains many other solutions. In particular,
we wish to investigate solutions which are globally spherically symmetric. The
general situation we wish to consider consists of a spherically symmetric
initia
   l
data set which has a lump of excess matter surrounded by a constant density
background which fills the rest of the universe. Outside the lump, the
Hamiltonian constraint guarantees that the scalar curvature is constant. The
combination of spherical symmetry and constancy of the scalar curvature
guarantees that that section of the manifold can be regarded as part of a
standard round $S^3$ [7]. This is {\it not} equivalent to guaranteeing that
the conformal factor must be identically one in the exterior region. We need
to understand the conformal structure of spherically symmetric constant
scalar curvature manifolds.

 Let us return to the Hamiltonian constraint, Eq. (19)
 $$\nabla^2\phi = {3 \over 4a^2}(\phi - \phi^5) - [{1 \over 12}K^2 +
2\pi(\rho - \rho_0)]\phi^5 \;, \eqno(75) $$
and let us consider the simplest possible case. Let us assume that the matter
is at rest, which gives us K = 0, and that the energy density is a constant,
$\rho =\rho_0$. The Hamiltonian constraint now simplifies enormously, to give
$$\nabla^2\phi = {3 \over 4a^2}(\phi - \phi^5). \eqno(76)$$
Obviously, one solution of (76) is $\phi \equiv 1$, which is the
standard solution. This is not the only one. There is a complete family of
regular solutions of Eq. (76) given by $$ \phi_{(\alpha)} = {\alpha^{1 \over 2}
\over (\alpha^2\cos^2{r \over 2} + \sin^2{r \over 2})^{1 \over 2}} \;,
\eqno(77)$$ where $\alpha$ is an arbitrary positive constant.

In Eq. (76) we are trying to find a conformal factor $\phi$ which transforms a
manifold with constant scalar curvature $^{(3)}R = 6/a^2$ to another manifold
with the same constant curvature  $^{(3)}R = 6/a^2$. The Yamabe Theorem
[8] states that every Riemannian manifold can be conformally transformed into
on
   e
with constant scalar curvature. The existence part of the Yamabe theorem has
bee
   n
shown to be true. Whether the resulting constant scalar curvature manifold is
unique is an open problem. However, it is known that in the case of flat space
we do not have uniqueness; there is at least a one parameter family
of inequivalent manifolds. The conformal factors mapping between them are just
the $ \phi_{(\alpha)}$'s given by Eq. (77). The manifold with metric
$$ d\-S^2 = \phi_{(\alpha)}^4 a^2[dr^2 + \sin^2rd\Omega^2] \eqno(78)$$
can be shown to be a round metric by using the coordinate transformation [7]
$$L = \pi - 2\tan^{-1}(\alpha\cot{r \over 2}) \;. \eqno(79)$$

It is important that we understand the general structure of the
$ \phi_{(\alpha)}$'s. It is clear from (77) that $ \phi_{(\alpha)}$ with
$\alpha = 1$ satisfies $\phi \equiv 1$ and so represents the identity
transformation. Each $ \phi_{(\alpha)}$, with $\alpha \not \equiv 1$, equals
$1/\sqrt{\alpha}$ at r = 0; equals $[2\alpha/(\alpha^2 + 1)]^{1/2}$, which is
always less than 1, at $r = \pi/2$; and equals $\sqrt{\alpha}$ at $r = \pi$.
If $\alpha < 1$ then $ \phi_{(\alpha)}$ monotonically decreases, if $\alpha
> 1$, $ \phi_{(\alpha)}$ monotonically increases. The value of r, call it
$R_1$, at which each $ \phi_{(\alpha)}$ passes through 1 is given by
$$\alpha = \tan^2{R_1 \over 2} \;. \eqno(80)$$

In the range $0 \leq r \leq \pi/2$ the $ \phi_{(\alpha)}$'s with $\alpha > 1$
give a smooth set of non-intersecting curves with one curve passing through
each point in the $(r, \phi)$ plane with $ \phi_{(\alpha)}$ in the range $ 0
< \phi < 1$. At each fixed value of r as $\alpha$ increases, $\phi$
decreases. Each individual curve is increasing as r increases but they are
all still less than 1 on reaching $r = \pi/2$. Beyond $r = \pi/2$ each curve
in turn crosses the $\phi = 1$ line at increasing values of $R_1$ with
increasing $\alpha$ (as given by (80)) without intersecting any other curve.
However, each $ \phi_{(\alpha)}$ is climbing more rapidly than the curves
with lower values of $\alpha$  by the time it crosss the line $\phi = 1$.
Above this line the curves proceed to cross because a curve with higher
values of $\alpha$ must rise above all the curves with lower $\alpha$ by the
time it reaches $r = \pi$.

Superimposed on this pattern is a mirror-image set of curves for values of
$\alpha < 1$. These start with values greater than 1 at r = 0, the smaller
$\alpha$ the greater the value. They then proceed to decrease as r
increases, crossing one another as the curves with smaller values of
$\alpha$ fall below the curves with larger values of $\alpha$. All the
crossing is accomplished by the time they cross the $\phi = 1$ line, and
this all happens before one reaches $r = \pi/2$. Beyond $r = \pi/2$ all the
curves smoothly decrease without crossing until one reaches $r = \pi$. The
part of $(\phi, r)$ space defined by $\phi^2 \sin r > 1$ contains no
curves. Through every other point two curves, with different values of
$\alpha$, pass.

Now let us consider one of these constant scalar curvature manifolds, with
line-element given by (75).When we substitute the explicit form of
$\phi_{(\alpha)}$ into Eq. (21) we get
 $$\nabla_an^a= 2(a\alpha\sin
r)^{-1}[\alpha^2 \cos^2 (r/2) - \sin^2 (r/2)] \;. \eqno(81)$$
 This quantity has only one zero at $r = R_E$ given by
$$\alpha = \tan {R_E \over 2} \;. \eqno(82)$$
An alternative form is
$$\cos R_E = {1 - \alpha^2 \over 1 + \alpha^2} \;. \eqno(83)$$
This value ($R_E$), from (79), corresponds to $L = \pi/2$, i.e., the physical
equator of the conformally transformed space. $\nabla_an^a$ is positive for all
$r < R_E$ and is increasingly negative for $r > R_E$. Therefore, independent of
the value of $\alpha$, all the manifolds share the property of the standard
Friedmann slice that they have no trapped surfaces in the `northern'
hemisphere,
but that as one approaches the `south' pole all the two-spheres become trapped.

The equator, where $\alpha = \tan {R_E \over 2}$, also plays another role
which will be important in our future discussions. Consider one of these
curves $ \phi_{(\alpha)}$, with a given $\alpha$, and let us assume
$\alpha > 1$. Start on this curve at r = 0, with $\phi = 1/\sqrt{\alpha} < 1$.
As one moves along this given curve, other  $\phi_{(\alpha)}$'s, with
different values of $\alpha$, cross it. These curves will have $\alpha < 1$.
These begin with curves whose value of $\alpha$ is essentially zero. As r
increases, as we move along our chosen curve, the value of $\alpha$ of the
curves crossing it monotonically increases, but is still less than 1 as we pass
$r = \pi/2$. Finally, at $R_1$, given by Eq. (80), with $R_1 > \pi /2$, the
curv
   e
crosses the line $\phi = 1$ (which is, of course, the  $\phi_{(\alpha)}$ with
$\alpha = 1$). The chosen curve now starts to overtake curves with values of
$\alpha > 1$ but less than our chosen $\alpha$. This continues until one
reaches
the the equator, $r = R_E$, given by eqns.(79) and (80). Note $R_E > R_1$
for $\alpha >1$. This point also satisfies $\phi^2\sin R_E = 1$ so is the
boundary of the excluded region. Only one curve, our chosen one, passes through
this point. From this point on , as we move along our curve in a direction of
increasing $r$, our chosen curve is now being overtaken by curves
$\phi_{(\alpha)}$ with larger and larger $\alpha$'s. Just as the vertical line
$r = 0$ is essentially the curve $\phi_{\alpha}$ with $\alpha = 0$, the
vertical line $r = \pi$ is the $\phi_{\alpha}$ with $\alpha = \infty$.

 \vskip 0.5cm
\centerline{\bf X Spherically symmetric models with non-constant scalar
curvature}
\medskip
We wish to construct a manifold which is spherically symmetric with a
given (non-constant) scalar curvature R. We can assume that the metric
is of the form (9), and then we can write the equation satisfied by $\phi$
in a form very similar to equation (75), i.e.,
$$\nabla^2\phi = {3 \over 4a^2}(\phi - \phi^5) - {1 \over 8}\Delta R\phi^5 \;,
    \eqno(84)$$
where $\Delta R = R  - 6/a^2$.

As we mentioned earlier, we wish to consider manifolds with constant density
outside some compact region. Thus we will wish to consider situations where
$\Delta R$ is zero outside some finite spherical volume. In the exterior region
the equation for $\phi$ reduces to
 $$\nabla^2\phi = {3 \over 4a^2}(\phi - \phi^5). \eqno(85)$$
 We know a complete family of solutions to this equation, the
$\phi_{(\alpha)}$'s of Eq. (77). However, it is not immediately obvious that
the
solution of Eq. (85), which is only valid on part of the manifold, must be one
of these $\phi_{(\alpha)}$'s. In one case we need regularity both at $r = 0$
and
at $r = \pi$. In the other case we need regularity only at, say, $r = \pi$. We
know that any solution to (84) must have some finite positive value at $r =
\pi$, and that it's first derivative must vanish there. We have a
$\phi_{(\alpha)}$ with the same value at $r = \pi$ and with the same first
derivative. This would be usually enough to prove that the two functions must
coincide. We cannot immediately deduce this here because the point $r = \pi$ is
a singular point of the equation because $\sin r = 0$ there. Happily, Eq. (85)
i
   s
really only an ordinary differential equation because of the spherical
symmetry and a treatment exists of such singular points in Rendall and
Schmidt [9]. Using their Theorem 1 allow us to deduce that the two solutions
must coincide and that the solution of Eq. (85) must be one of
the $\phi_{(\alpha)}$'s in the exterior of the support of $\Delta R$.

 One obvious situation we would like
to consider is when there is a step-function in the scalar curvature, when
the scalar curvature has one (constant) value in part of the manifold and
has a different (constant) value on the rest. It is easy to see that the
conformal factor to achieve this must made up of a pair of the functions
$\phi_{(\alpha)}$ defined by Eq. (77). More precisely, we would want that
the conformal factor equal $C_1\phi_{(\alpha_1)}$ in one region and
$C_2\phi_{(\alpha_2)}$ in the rest, where $C_1$ and $C_2$ are prescribed
constants depending on the values of the scalar curvature and $\alpha_1$
and $\alpha_2$ are adjustable parameters. One now tries to match these
functions and their first derivatives at some radius and discovers that
this can not be done, irrespective of the values of $C_1$, $C_2$ or the
coordinate value of the matching point..

This means that we cannot expect to be able to solve Eq. (84) with a
randomly chosen $\Delta R$. Equally, if we get a solution it may well
not be unique. Nevertheless, we have been able to extract a number of
interesting and useful properties of solutions to Eq. (84) (assuming one
exists!). The situation we are interested in is when $\Delta R \geq 0$ and
has compact support. In the exterior region we know that the solution
must be one of the $\phi_{(\alpha)}$ 's. We will show that this $\alpha$
must satisfy $\alpha < 1$ if the support of $\Delta R$ is not too large.

The assumption we will make is that the support of $\Delta R$
lies entirely in the `northern' hemisphere. We do not define this in
terms of the background geometry, but rather in terms of the physical
metric. The exterior solution is defined by a $\phi_{(\alpha)}$ with
$\alpha = \alpha_0$. We assume that the `equator', the radius $R_E$
satisfying
 $$\alpha_0 = \tan R_E/2 \;, \eqno(86)$$
 lies in the exterior zone.

 The analysis is based entirely on the maximum principle as applied to
Eq. (84). The solution $\phi$ to (84) cuts through a complete
family of $\phi_{(\alpha)}$'s. At any point along $\phi$, we know that the
two $\phi_{(\alpha)}$'s that pass through that point satisfy
$$\nabla^2\phi_{(\alpha)} = {3 \over 4a^2}(\phi_{(\alpha)} - \phi_{(\alpha)}^5)
\;. \eqno(87)$$ Therefore we must have at that point since we assume $\Delta R
\geq 0$ $$\nabla^2(\phi_{(\alpha)} - \phi) \geq 0 \;.\eqno(88)$$
If $\Delta R > 0$ at this point then by continuity $\nabla^2(\phi_{(\alpha)} -
\phi) > 0$ close to this point. This implies, due to the maximum principle,
that
if $\phi$ approaches a $\phi_{(\alpha)}$ from above it must pass through it; it
cannot either `bounce' off it or merge with it.

 We obtain our result by contradiction, so we begin by assuming that
$\alpha_0 > 1$. Let us start at $r = \pi$ and move along the curve $\phi$,
the solution to (84). In this region it coincides with
 $\phi_{\alpha_0}$. We continue along this curve through $R_E$ (as defined
by (85)). At some radius $ r < R_E$ we enter the support of $\Delta R$. At
this point the curve $\phi$ must bend downwards, away from
$\phi_{\alpha_0}$. But in this region the space underneath
$\phi_{\alpha_0}$ is filled with all the curves  $\phi_{(\alpha)}$ with
$\alpha > \alpha_0$. Our solution curve must continue descending faster
than these  $\phi_{(\alpha)}$'s. As we point out above, it can neither
merge with, nor bounce off, any of them. This can only lead to disaster.
Either $\phi$ goes to zero, or hits the r = 0 point with a non-zero angle
(if $\phi' = 0$ at r = 0, it would be tangent to one of the
$\phi_{(\alpha)}$'s). Thus the original assumption, that $\alpha_0 > 1$
cannot be true. A similar disaster befalls a curve which starts with
$\alpha_0 = 1$. Therefore we must conclude that $\alpha_0 < 1$.

Each point of $(r, \phi)$ space (excepting the region defined by $\phi^2\sin r
\ge 1)$ has two curves $\phi_{(\alpha)}$ passing through it. Thus any curve
$\phi(r)$ can be described by giving the pair of $\alpha$ values corresponding
to each point it passes through. Consider the particular curve we describe in
the preceeding paragraph. As it moves away from $r = \pi$ one of the $\alpha$'s
remains fixed at $\alpha_0$, the other starts at $\infty$ and monotonically
decreases, at $R_E$ both $\alpha$'s coincide, and at values of $r < R_E$ the
varying $\alpha$ is less than $\alpha_0$. Now we enter the support of $\Delta
R$. The curve $\phi(r)$ now drops below the curve $\phi_{(\alpha_0)}$. One
$\alpha$ continues to decrease, the other increases above $\alpha_0$. The key
question is what happens to this increasing $\alpha$ ? It cannot reach a
maximum value either at or before reaching $r = 0$, because this is equivalent
to being a tangent (from above) to one of the $\phi_{(\alpha)}$'s, which is
forbidden by the maximum principle.
Thus something bad must happen to the curve in question, it either goes to zero
at a nonzero value of $r$, or $\phi' > 0$ at $r = 0$.

 We can now deduce somewhat more assuming, as always, that $\Delta R
\geq 0$ and that it has support only on one side of the equator. We now
assume $\alpha_0 < 1$. Let us follow the solution to (84) in from $r = \pi$.
We start off at a value of  $\phi = (\alpha_0)^{1/2} < 1$ and it monotonically
increases as we move along
  $\phi_{\alpha_0}$. We will not deviate from this curve until we, at
least, reach the equator ($R_E$). By this point  $\phi_{\alpha_0}$ has
risen above 1. The solution $\phi$ starts curling downwards  relative to
 $\phi_{\alpha_0}$ at some $r < R_E$. However, it cannot curl down too
much, because if it passes, going downwards, through $\phi = 1$ it is
stuck in the same trap as before. We therefore can assume that $\phi > 1$
in the interior. But , as long as $\phi$ is $> 1$, we have $\nabla^2\phi < 0$
and this means that $\phi$ cannot have a minimum in the support of $\Delta
R$. Hence we must have $\phi' \leq 0$ in the interior, with equality only
at r = 0. Of course, since $\alpha_0 < 0$ we have $\phi' < 0$ in the
exterior.

The result that the conformal factor outside the support of the excess matter
must coincide with one of the special functions $\phi_{(\alpha)}$ is
interesting insofar as it shows that the exterior region is homogeneous and
isotropic, uninfluenced by the lump. This is exactly analogous with the result
derived in Section II that the current potential K vanishes outside the support
of the current.

Let us now return to the promised proof of Lemma 6 from Section VIII. Here we
consider a situation where part of the manifold consists of a region $(0, r_0)$
near the north pole with $\Delta R \ge 0$, a second region $(r_0, r_1)$  with
$\Delta R \le 0$ and that we have standard FLW data outside these regions.
This means that the conformal factor, the solution to (20), equals 1 outside
$r_1$. If we follow the conformal factor $\phi$ in from the right it must
increase as it enters the region of negative $\Delta \rho$. If it went down
there would be an interval $(r_1 - \delta, r_1 + \delta')$
in which $\phi \le 1$. But we have from (83), since $\Delta R \le 0$,
$$\nabla^2 \phi \ge 0 \;.$$
However $\phi$ achieves an interior maximum at $r = r_1$, which is forbidden.
Therefore we must have $\phi \ge 1$ as we approach $r_1$.

Having shown that $\phi > 1, \phi' < 0$ near $r_1$, we now wish to show that
$\phi > 1$ in the whole interval $(r_0, r_1)$. Let us assume that $\phi' = 0$
somewhere in the interval $(r_0, r_1)$. If $\phi^2 \sin r < 1$ in the interval,
this means that $\phi$ must turn over in the region that is filled with
$\phi_{\alpha}$ curves. In particular, this means that the $\phi$ curve must
tangent from below one of these curves, say $\phi_{\alpha_0}$, at some point,
say $r = r_3$. At this point we have
$$\phi(r_3) = \phi_{\alpha_0}(r_3) > 1,~~~\phi'(r_3) = \phi_{\alpha_0}(r_3) <
0 \;.\eqno(89)$$
In a neighbourhood of $r_3$ we also have
$$\phi \leq \phi_{\alpha_0} \;. \eqno(90)$$
Subtracting (87) from (84) gives
$$\nabla^2(\phi_{\alpha_0} - \phi) = {3 \over 4a^2}[\phi_{\alpha_0} -
\phi^5_{\alpha_0} - \phi + \phi^5] + {1 \over 8}\Delta R \phi^5 \;. \eqno(91)$$
The right-hand-side of (88) is non-positive, so the minimum principle says that
$\phi_{\alpha_0} - \phi$ cannot achieve a minimum at $r_3$, which contradicts
our assumptions. Hence $\phi' \leq 0$ on the whole interval $(r_0, r_1)$, with
equality only in the FLW region. Therefore $\phi(r_0) > 1$. Now we can use
Lemma 5 to finally show $\phi > 1, \phi' < 0$ in the interval $(0, r_0)$.

\vskip 0.5cm
\centerline{\bf XI Trapped surfaces in spherical universes.}
\medskip
In the previous section we have proven that a spherical bump
of  compact support
confined to the northern hemisphere produces a monotonically
decreasing conformal factor $\phi$,
$$\phi ' \le 0 \;,$$
if its excess energy is non-negative.
This  corresponds to Lemma 5 of Sec. VIII. We can also prove
$$\phi'\sin r + \phi \cos r \ge 0$$
Outside the bump $\phi $
must coincide with $\phi_{\alpha }$, where $\alpha < 1$. As is shown
in Lemma 7, the quantity  $\phi' \sin r + \phi \cos r$ monotonically
decreases. Outside the matter it is equal to
$$\alpha ^{-1} \phi^3 [(1 + \cos^2r)(\alpha^2-1) + 2\cos r (1+\alpha^2)] \;.$$
This expression vanishes at a value $R_1$ such that $\cos R_1 =
(1-\alpha)/(1+\alpha )$, (Eq. (79)),i.e., when $\phi_{\alpha }$ crosses through
1. Therefore inside a region with non-negative matter density the
inequality (73) holds.

Thus we can prove:
\medskip
{\bf Theorem III:} Given a globally spherically symmetric initial data set
which satisfies trK = constant with a localized positive lump which is
instantaneously at rest.
If we can find a spherical surface satisfying
$$\Delta M >  L + {AH \over
4\pi} -{3V \over 8 \pi a^2} $$
then that surface is trapped.
\medskip
We can also prove:
\medskip
{\bf Theorem IV:} Given a globally spherically symmetric initial data set
which satisfies trK = constant with a localized positive lump may be moving.
 If we can find a spherical surface satisfying
$$\Delta M - \Delta J>  {7 \over 6}L + {AH \over
4\pi} -{3V \over 8 \pi a^2} $$
then that surface is trapped.
\medskip
 In addition, one can prove another version
of a sufficient condition (Lemma 3) for the formation of
trapped surfaces by moving inhomogeneities.
Indeed, let us return to the equation (68). In section VII we estimated
$\Gamma $ from above by $H L^2$.
The other estimate is that $\Gamma < 8Ha^2$. This is obtained when
one realises that $\phi$ merges with some $\phi_{\alpha}$, with $\alpha < 1$ in
the exterior, and that it must lie below this $\phi_{\alpha}$ in the
interior. In turn, we know $\phi_{\alpha} < \alpha ^{-1/2}$. Thus, if we
replace $\phi^4$ in (82) with $\alpha^{-2}$ we will get an upper bound
$$\Gamma < 4Ha^2 \alpha^{-2}\int_0^R\sin r \cos r dr = 2Ha^2 \alpha^{-2}
\sin^2 R \;. \eqno(92).$$
This, as it stands is not a particularly interesting inequality until one
realises that we are assuming that the excess matter is confined within the
equator and that we are not interested in finding trapped surfaces beyond
the equator. Therefore, we can replace $R$ in (92) by $R_E$ as given by
(83). It is a straightforward manipulation of trigonometric functions to show
$$\alpha^{-2} \sin^2 R_E = {4 \over (1 + \alpha^2)^2} < 4 \;, \eqno(93)$$
which gives the desired inequality.
Using (89) and proceeding as in Section VI, we obtain the following
version of Lemma 3:
\medskip
{\bf Theorem V}: Under the preceeding conditions, if
we obtain a surface for which
$$\Delta M - \Delta J > {7 \over 6}L + {2a^2H \over \pi} - {3V \over 8\pi a^2}
\;, \eqno(94)$$
either the surface itself is trapped or the interior contains trapped
surfaces.
\vskip 0.5cm
\centerline{\bf XI Concluding remarks.}
\medskip
Let us summarize the whole discussion.
In this paper we study spherically symmetric closed universes. There are
two different situations which are of interest.

First, we may have a homogeneous and spherically symmetric background
geometry with many spherical bumps placed in it. Globally, a resulting
universe is neither spherically symmetric nor homogeneous, but locally,
close to a particular lump, there is a spherical symmetry with respect
to the centre of the lump. Also, very far from the bump, the geometry of
the chosen Cauchy slice coincides with the  homogeneous
and spherically  background geometry. That case was studied in Section
VIII.
 In Section VIII is formulated a sufficient
condition (Theorem II) for the formation of trapped surfaces by $moving$
perturbations, when the initial momentum of the gravitational field
is changed. Obviously, Theorem II can be specialized to the case where the
excess matter is at rest. Theorem I in Section V gives us a nice necessary
condition for the formation of a trapped surface when the matter is at rest.
Interestingly, we have failed to find an equivalent necessary condition in the
case where the matter is moving. In Section VII we prove  that in a sphere of a
fixed radius only a finite amount of perturbed energy can be packed.

Secondly, one might be interested in investigating geometry of a universe
that is not homogeneous but is globally spherically symmetric.
Section IX contains a description of spherically symmetric constant scalar
curvature  models. These solutions are used in Section X to derive a number  of
properties of initial data  for nonhomogeneous distributions of matter. Using
them, we prove in Sec. XI criteria for the formation of trapped surfaces in
globally spherically symmetric geometries that are generated by a single bump
confined to only one  hemisphere.

We would like to stress that these results obtained here are correct in the
full nonlinear theory, nowhere do we draw on a perturbation expansion or a
linear approximation.  We expect that these results are  true also for a class
of  nonspherical perturbations, for instance in the class of conformally flat
perturbations of the homogeneous closed universes we would expect to obtain
results  analogous to those known in  open flat universes [10].
\par
{\bf Acknowledgements.}
We would like to thank the Physics Department, University College Cork
for inviting one (ME) of us to Cork, where  part of
this work was done. N. \'O  M. would like to thank the members of the Max
Planck Institut f\"ur Astrophysik in Garching, especially Alan Rendall and Uwe
Brauer, for their help and interest. N. \'O M. would also like to thank Jemal
Guven. This work has been  supported by the  Polish Government Grant no. PB
2526/2/92. \vfill\eject \centerline{\bf References.}
\medskip
[1] R. Penrose, {\it Phys. Rev. Lett.} {\bf 14}, 57 (1965);{\it Techniques
of Differential Topology in Relativity} (Soc. Ind. Appl. Math. 1972).\par
[2] S. W. Hawking, G.F.R. Ellis, {\it The large scale structure of
space-time} (Cambridge University Press, Cambridge 1973). \par
[3] C. W. Misner, K. Thorne, J. Wheeler, {\it Gravitation}
(Freeman, San Francisco 1973).\par
[4] P. Bizo\'n, E. Malec, N.\'O Murchadha,
{\it Phys. Rev. Lett.} {\bf
61}, 1147(1988); {\it Class. Quantum Grav.} {\bf 6}, 961 (1989); {\it Class.
Quantum Grav.} {\bf 7}, 1953 (1990).

[5] U. Brauer, E. Malec, {\it Phys. Rev.} {\bf D45}, R1836 (1992).

[6] A. Einstein {\it Ann. Math.} {\bf 40}, 922 (1939).

[7] J. Guven and N. \'O Murchadha, to be published.

[8] R. Schoen, {\it J. Diff. Geom.} {\bf 20}, 479 (1984).

[9] A. Rendall, B. Schmidt, {\it Class. Quantum Grav.} {\bf 8}, 985 (1991).

[10] P. Koc, E. Malec, {\it Acta Phys. Pol.}{\bf B23}, 123 (1992).

\end